\documentclass[12pt]{iopart}
\usepackage{graphicx}
\usepackage{bm}      
\usepackage{epsfig}
\usepackage{wasysym}
\usepackage{cases}
\usepackage{fullpage}
\usepackage{setspace}
\usepackage{bm}
\usepackage{color}
\begin{document}

\title{The valley filter efficiency of monolayer graphene and bilayer graphene line defect model}

\author{Shu-guang Cheng,$^{1}$ Jiaojiao Zhou,$^{2}$ Hua Jiang,$^{2}$\footnote[1]{Author to whom any correspondence should be addressed.} Qing-Feng Sun$^{3,4}$}

\address{$^1$Department of Physics, Northwest University, Xi'an 710069, People's Republic of China\\
$^2$College of Physics, Optoelectronics and Energy, Soochow University, Suzhou, 215006, China\\
$^3$International Center for Quantum Materials, School of Physics, Peking University, Beijing 100871, China\\
$^4$Collaborative Innovation Center of Quantum Matter, Beijing 100871, China}
\ead{jianghuaphy@suda.edu.cn}

\begin{abstract}
In addition to electron charge and spin, novel materials host another degree of freedom, the valley. For a junction composed of valley filter sandwiched by two normal terminals, we focus on the valley efficiency under disorder with two valley filter models based on monolayer and bilayer graphene. Applying the transfer matrix method, valley resolved transmission coefficients are obtained.
We find that: i) under weak disorder, when the line defect length is over about $15\rm nm$, it functions as a perfect channel (quantized conductance) and valley filter (totally polarized);
ii) in the diffusive regime, combination effects of backscattering and bulk states assisted intervalley transmission enhance the conductance and suppress the valley polarization;
iii) for very long line defect, though the conductance is small, polarization is indifferent to length. Under perpendicular magnetics field, the characters of charge and valley transport are only slightly affected. Finally we discuss the efficiency of transport valley polarized current in  a hybrid system.
\end{abstract}

\section{INTRODUCTION}

Electron manipulation in solid state physics reveals the nature of materials and provides potential application in nano-devices. In addition to electron charge, electron spin controlling inspires the field of spintronics.\cite{GMR1,spintronics1,spintronics2,spintronics3} Relying on spin-orbital coupling, spin Hall effect\cite{spinHall} and quantum spin Hall effect\cite{QspinHall1,QspinHall2} drive electrons of different spins deflect in opposite directions, in analogy to electrons and holes in quantum Hall effect.\cite{QHall} Novel two dimensional honeycomb lattice materials, such as graphene\cite{Graphene}, silicence\cite{Silicene} and transition metal dichalcogenides,\cite{MoS2A} host valley degree of freedom. So the manipulation of valley degree of freedom for electrons in these material breeds the new field of valleytronics.\cite{Valley1,Valley3,Valley4,Valley5,ValleyHall1, ValleyHall2,ValleyHall3,ValleyHall4,Valleyline}

Valley filters, serving as valley polarized current generator and allowing only carriers of a specific valley passing through, are one important type of valleytronics devices.\cite{Beenakker} Theoretical researches predicted that kink states in the line defect of few layer graphene host the transmission of valley polarized current.\cite{Domain1,Domain5,Domain2,Domain3, add1,Domain4,Domain6,Filter,add2} In a hexagonal lattice, when the effective mass of carriers changes its sign across the line defect, the kink states are formed along the line defect with zero mass. Significantly, the propagation direction of the kink states is related to the valley index. In other words, carriers belong to different valleys propagate in opposite directions. So a clean line defect serves as a valley filter, in analogy to spin filter in spintronics. Besides, the performance of the valley filter can be controlled electrically. In monolayer graphene, the line defect can be induced by aligning the monolayer graphene sample differently to hexagonal boron nitride substrate at two sides of the line defect. In detail, in one side, carbon atoms in sublattice $\rm A$ have higher on-site energy than carbon atoms in sublattice $\rm B$ while in the other side, the case reverses.\cite{Domain1,add1,Domain6} For bilayer graphene, the line defect can be realized in two ways. i) One can apply an electric field perpendicular to the sample while in two sides of the line the direction reversed.\cite{Domain5,add2,DomainBE1} ii) Generating a line dislocation between two layers thus on the left (right) side of the line, the bilayer graphene is $\rm AB$ ($\rm BA$) staggered. Then a uniform perpendicular electric field is applied to break the inversion symmetry.\cite{DomainBE2,DomainBE3,DomainBT1,DomainBT2} Above two models are in physics equivalent. Based on the models, very recently experiments confirm the existence of kink states by electron transport \cite{DomainBE1} and scanning tunneling microscope \cite{DomainBE3}, respectively. Although theoretical results indicate that the linear conductance through a line defect in bilayer graphene should be $2G_0$ ($G_0=2e^2/h$ the quantum conductance), the experiment detected value is smaller, indicating that the ideal value is affected by disorder and the limited sample size.

The efficiency of a valley filter is the most important index, dictating the polarization of the output current. Thus the manipulation of valley efficiency is the primary task of valley filter study. First, valley index is a degree of freedom based on momentum and is highly related to the translation invariance of crystal lattice. So for samples of limited size or dirty samples with disorder, translation invariance is broken and momentum is not a good quantum number. So we wonder under such circumstances, whether the valley filter still perform high valley efficiency. Second, when valley filters are connected for applications, the connecting material is usually different from that of valley filter. The performance of a valley filter in such case is not clear yet.

In this paper, we investigate the generation of valley resolved kink states in line junctions based on monolayer and bilayer graphene models. By using the transfer matrix method on lattice models, the intervalley and intravalley transmission coefficients, hence the total conductance and valley polarization efficient are obtained. In the clean limit, when the length of the line defect is over $15 \rm nm$, both models serves well as valley filter with quantized conductances $G$ ($G=G_0$ for monolayer graphene model and $G=2G_0$ for bilayer graphene model without spin degeneracy) and nearly total valley polarized. In the diffusive regime, the transmission is affected by the combination of backscattering and bulk states assisted intervalley transmission. Consequently, $G$ is enhanced and valley polarization is suppressed. In very strong disorder, all transmissions are destroyed that both conductance $G$ and valley polarization
$P$ decreases fast. We focus on the effects of disorder type (Anderson type and long range disorder) and magnetic field on the valley resolved transport.
The valley resolved transport in hybrid system is also discussed.

In section \ref{model} we give the tight-binding models and related parameters used in numerical calculation. The main results are given in section \ref{results}, including i) how long a line defect is needed to function as a perfect valley filter; ii) the performances of monolayer and bilayer graphene model under disorder; iii) the performance of valley filter for different length; iv) the efficiency of injecting valley polarized current into normal monolayer graphene. Finally a brief conclusion is given in section \ref{conclusion}.

\section{Model and methods}\label{model}

\begin{figure}
\begin{center}
\includegraphics[width=\columnwidth, viewport=60 80 630 560, clip]{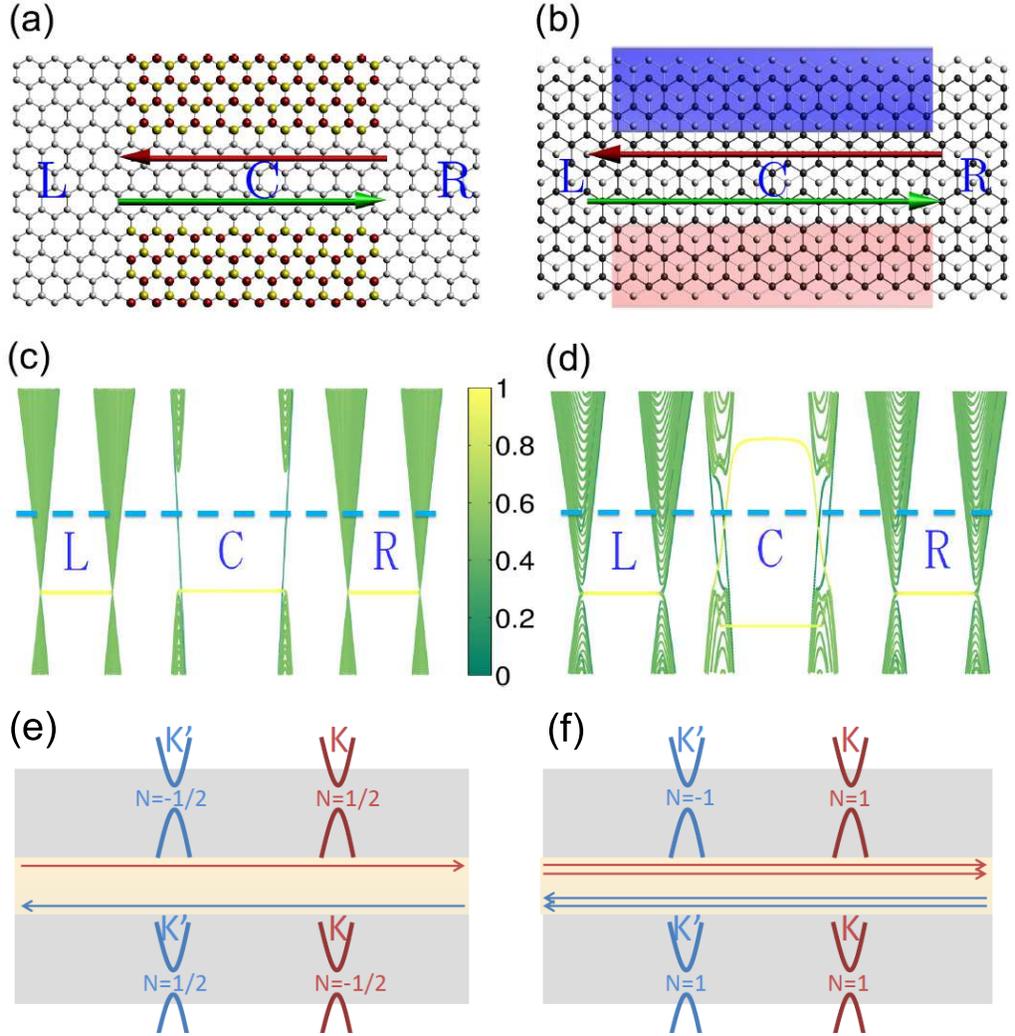}
\caption{(color online) The schematics of (a) monolayer and (b) bilayer graphene based valley filters. Letters $\rm L$, $\rm C$ and $\rm R$ denote the left terminal, central region and the right terminal. The size parameters are (a): the width of the ribbon $N=12$, the length of the line defect $M=9.5$ and the width of the domain wall $d=4$; (b) $N=9$, $M=10$ and $d=4$. The green (grey)/brown (black) arrow indicates the propagation direction of mode with valley index $K$/($K'$). In (a), brown (black)/yellow (grey) balls in region $\rm C$ represent carbon atoms with staggered on-site energy $\delta$/($-\delta$) and light grey balls symbol pristine atoms. In (b), black and grey balls in region $\rm C$ represent the top layer and bottom layer of bilayer graphene, respectively. The blue (heavy grey) and red (light grey) rectangular areas in (b) represent gate voltage applied in substrates for the purpose of inversion symmetry breaking.
The band structure of model (a) is shown in (c) with ribbon width $N=160$. The band structure of model (b) is shown in (d) with width $N=120$ in region $\rm L, R$ and $N=180$ in $\rm C$. Region $\rm C$ in (b) is wider than $\rm L$ and $\rm R$ so the transmission through edge states is annihilated. Under such parameters, the band gaps in two models are the same. The dashed vertical lines are the Fermi level $E$ and the colors of the band reveal the position weighted average of the eigenstates. So in (c)/(d), for each valley, there are one /two kink states cross the Fermi level. The degenerate states within the gap in (d) cross the Fermi level with yellow (light grey) color describe the edge states. The schematic diagrams of the topological charges for monolayer (e) and bilayer (f) graphene line defect models.
}\label{Fig1}
\end{center}
\end{figure}

The model we mainly focused (except subsection \ref{PartV}) is constructed by a line junction sandwiched by two terminals as displayed in Fig. \ref{Fig1}(a) or (b) with $\rm L$, $\rm C$ and $\rm R$ indicate the left terminal, central region and right terminal. Left and right terminals serve as normal current source and drain, respectively. The central region is the valley filter with a line defect in middle. For the central region long enough, electrons propagate in opposite directions belong to different valleys. So one expect that the normal electrons (not valley polarized) injected from left terminal, after propagating along the line junction, flow into the right terminal with large valley polarization. Here we consider two types of models, one is based on the monolayer graphene and the other is based on the bilayer graphene model. The Hamiltonian of the monolayer graphene models in the tight-binding representation is,\cite{Domain1}
\begin{equation}\label{EQ1}
H_s=\sum_{i} (\epsilon_i+w_i) c^{\dagger}_ic_i+\sum_{ \langle ij \rangle} t (e^{i\phi_{ij}}c^{\dagger}_i c_j+e^{-i\phi_{ij}}c^{\dagger}_j c_i),
\end{equation}
with $c_i$ the annihilation operator for electrons at site $i$ and $\epsilon_i$ is the on-site energy
at site $i$. The sum subscript $ \langle ij \rangle$ in the second term counting the nearest neighbor coupling with parameter $t$. The phase factor $\phi_{ij}$ means the phase contributed by vector potential accounting the perpendicular applied uniform magnetic field. $w_i$ is introduced to account
disorder which will be detailed later. To form a line defect, we need to change the on-site energy
of $A$ and $B$ sublattices in region $\rm C$. For example, in one side of the line, we set $\epsilon_A=-\epsilon_B=\delta$ while in the other side, $\epsilon_A=-\epsilon_B=-\delta$. This can be realized in experiment by aligning the monolayer graphene with a line defect to hexagonal boron nitride substrate.\cite{Domain6}
In two terminals $\rm L$ and $\rm R$, $\epsilon_i=V_g$ which can be tuned by $V_g$ the applied gate voltage in the experiment. $V_g$ is set large that there are dozens of states at the Fermi level to guarantee that the injected electrons are of small, if not zero, valley polarized (see Fig. \ref{Fig1} (c) and(d)).
Thus the injecting electrons are both from both $K$ and $K'$ valleys.
Furthermore, bias voltage is applied that net electrons flow from the left terminal to right.

For Bernal staggered bilayer graphene model,\cite{add2}
\begin{eqnarray}\label{EQ2}
H_b&=&\sum_{i\in T\& B} (\epsilon_i+w_i) c^{\dagger}_ic_i+\sum_{ \langle ij \rangle \in T\& B} t (e^{i\phi_{ij}}c^{\dagger}_i c_j+e^{-i\phi_{ij}}c^{\dagger}_jc_i)\nonumber \\& +&\sum_{i\in T; j\in B} t_{\perp}(c^\dagger_ic_j+c^\dagger_jc_i).
\end{eqnarray}
Here the first two terms are similar to Eq. (\ref{EQ1}) including two layers (the top layer and bottom layer) but with no staggering energy. In the region $\rm C$, instead of AB sublattice staggering energy in Eq. (\ref{EQ1}), the top and bottom layer have different on-site energy applied by the gate voltage. There is asymmetry top-bottom energy different between the line junction, in one side, the top layer have higher on-site energy ($\epsilon_{T}=-\epsilon_{B}=U$) and in the other side, the situation reverses ($\epsilon_{T}=-\epsilon_{B}=-U$). The third term is coupling between two nearest site of top layer and bottom layer. Similarly the Fermi level is shifted away from the neutral point to allow more propagation channels in the left and right terminals. For the sake of concise expression we use model I and II for monolayer graphene model and bilayer graphene model, respectively.

To investigate the effect of disorder, we introduce Anderson type disorder and long range disorder in the central region. For the former type, at each site, $w_i$ is uniformly distributed in the interval $[-W/2, W/2]$ with $W$ the disorder strength. To simulate the long range disorder, the on-site energy at site $i$ is $w_i={\sum_j \tilde{w_j}exp(-r^2_{ij}/2\xi^2 )}$ where $\tilde{w_j}$ is the uncorrelated Anderson type disorder and $r_{ij}$ is the distance between site $i$ and $j$. $\xi$ is the parameter describing the correlation length. In the numerical calculation, we set $\xi=1.6a$ with $a$ the distance of nearest carbon atoms and the long range disorder with density $1\%$. With each value of disorder strength, the transmission coefficients are averaged upto 1000 configurations.

Next we use the transfer matrix method to calculate the transmission coefficients.\cite{Ando} For instance, there are $m$ ($n$) propagation mode in the left (right) terminal labeled by $m_i$ ($n_j$). The transmission coefficient from mode $m_i$ in terminal $\rm L$, through the scattering region $\rm C$, to mode $n_j$ in terminal $\rm R$, $T_{n_jm_i}$ is obtained. With no consideration of spin degeneracy, the transmission coefficient from valley $K_1$ in terminal $\rm L$ to valley $K_2$ in terminal $\rm R$ is
\begin{equation}\label{EQ3}
T_{K_2K_1}=\sum_{n_j\in K_2 \&m_i\in K_1}T_{n_j m_i},
\end{equation}
with $K_1, K_2\in K, K'$.
The linear conductance of the junction is the summation of Eq. (\ref{EQ3}): $G(E)= (e^2/h) \sum_{K_1,K_2}T_{K_2K_1}$. To investigate the valley polarization of transmission current, the valley efficient of a valley filter (e.g. valley $K$ is favored) is defined as
\begin{equation}\label{EQ4}
P(E)=(T_{KK}+T_{KK'}-T_{K'K}-T_{K'K'})/\sum_{K_1,K_2}T_{K_2K_1}.
\end{equation}

Before exhibiting the numerical results we detail the parameters in calculation. Throughout this paper, $t$ in Eq. \ref{EQ1} and \ref{EQ2} is chosen as the energy unit. In both models we consider zigzag edged nanoribbon for the purpose of valley related calculations. Here and hereafter, we normalize all energy scale by the hopping energy $t$ and in model II $t_{\perp}=0.15$. $\epsilon_i=0$ is set as the zero energy reference. To generate valley band structure in the central region, in model I the staggering energy is $\delta=0.05$ and in model II $U=0.1$. The values of $\delta$ and $U$ are such selected that band gaps of two models are all approximately $0.122t$ (see Fig. \ref{Fig1} (c) and (d)) which make the comparison under same disorder reasonable. For a uniformly staggered monolayer graphene or bilayer graphene with uniform electric field applied, there is energy gap determined by the magnitude of $\delta$ or $U$. The Chern number for valley $K$ and $K'$ are opposite to each other, $\tau_K=-\tau_{K'}$. For uniformly gapped monolayer (bilayer) graphene, $\tau_K=1/2 (1). $ As $\delta$ or $U$ changes its sign, $\tau_K$ becomes $-\tau_K$ and $\tau_{K'}$ becomes $-\tau_{K'}$. The number of channels per valley is determined by the Chern number difference across the line defect.\cite{Domain4}  Thus for each valley there are two channels for model I and four channels for model II with spin degeneracy considered. In Fig.\ref{Fig1} (e) and (f), the schematic diagrams of the topological charges are displayed for monolayer and bilayer graphene line defect models, respectively.

In experiment the line junctions are actually not `one-dimensional', or in other words, across the line defect the effective mass does not change abruptly. So we introduce a relaxation region to the middle line. For model I, along the middle line there is a region of width $d$ with no $\rm AB$ staggering energy. For model II, the energy difference between up and bottom layer across the line varies smoothly from positive to negative in a cosine way (from $1$ to $-1$) with width $d$.
What is more, except the kink states along the middle line, for model II there are also edge states. It can be seen from Fig. \ref{Fig1}(d) that there are two edge states across the Fermi level with lighter color (located at the edges of the ribbon).\cite{Biedge} So we need to eliminate the contribution of edge states in simulation. To do this, we set the width of the central region much wider than the terminals that all injecting modes will not be able to transport through the edge sates. In the whole discussion we use $N=160$ for model I. We set the width of the terminal $N=120$ and the width of the central region $N=180$ in model II. The size $N$ is wide enough to host one-dimensional kink states. In the presence of uniform magnetic field, $\phi=\frac{e}{2\hbar}\oint \textbf{A}\cdot d\textbf{l}=\frac{e}{2\hbar} \textbf{B}\cdot \textbf{S}$ with $\textbf{B}$ the magnetic flux density and $S$ the area of a single hexagonal lattice. So $\phi=0.001$ roughly equals to $25T$.

\begin{figure}
\includegraphics[width=\columnwidth, viewport=57 155 702 486, clip]{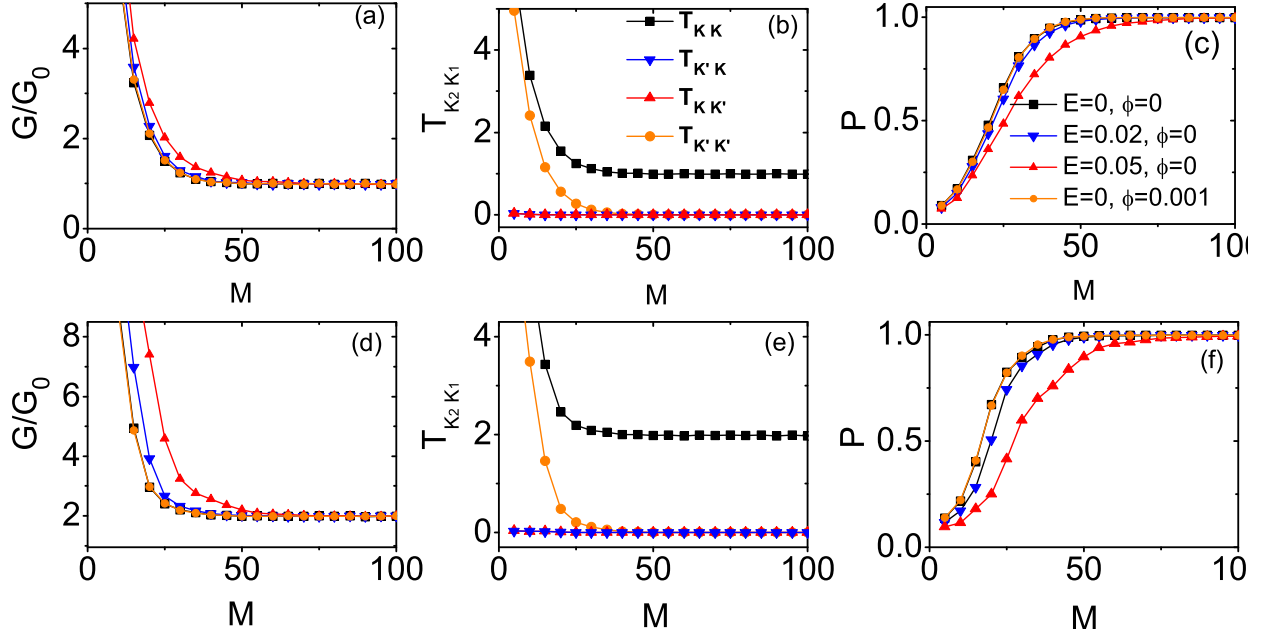}
\caption{(color online) The performance of valley filter as varying line defect length $M$ with upper panel for model I and lower panel for model II, respectively. (a) and (d): the linear conductance $G$ (in unit of $G_0$); (b) and (e): four valley resolved components of tunneling coefficients $T_{K_2K_1}$; (c) and (f): the valley polarization $P$ versus $M$.
}\label{Fig2}
\end{figure}

\section{NUMERICAL RESULTS AND DISCUSSION}\label{results}

We discuss the numerical results in this section. To discuss in a simple and clear way, for both models the right (left) going mode belong to valley $K (K')$. In fact it can be interchanged by reverse the sign of $\delta$ and $U$.

\subsection{Valley polarized current generation}\label{PartI}

First we focus on the valley polarized current generation. The first question is how long a line junction is needed to generate a current with sufficient high valley polarization. The results are shown in Fig. \ref{Fig2}. We inspect the behaviors of the conductance $G$, valley polarization $P$ and
valley resolved transmission coefficients $T_{K_2K_1}$ in several different situations, including varying Fermi level $E$ and magnetic field $\phi$.
Similar characters are found. For both models, as the length $M$ becomes large, the conductance $G$ decreases abruptly and saturates. For $M>60$ ($\sim 15 \rm nm$), $G$ is basically quantized ($G_0$ for model I and $2G_0$ for model II, see Fig. \ref{Fig2}(a) and (d)). Meanwhile, the valley polarization $P$ increases as $M$ becomes large and then saturates to $P=1$.
The valley resolved transmission coefficients $T_{K_2 K_1}$ are shown in Fig. \ref{Fig2}(b) and (e) for $d=20$, $E=0$ and $\phi=0$.
For both models, the intravalley transmission coefficient $T_{K'K'}$ decreases severely for large length $M$. When $M>40$, $T_{KK}$ saturates to quantized value ($G_0$ for model I and $2G_0$ for model II) and $T_{K'K'}$ decreases to zero. Meanwhile the intervalley transmission coefficients ($T_{KK'}$ and $T_{K'K}$) are comparatively small even for short $M$ and then decay to zero as $M$ becomes large. Above results indicate that for moderate long line defect, the line models work perfectly as valley filters. As the Fermi level $E$ approaches the band edge, the saturation of the conductance $G$ and valley polarization $P$ happens for larger length $M$. It is also confirmed that the performance of the valley filter is almost unaffected by $d$, the orientation of the straight line defect or even if the line defect is slightly curved (not shown). While the effect of magnetic field is quite neglectable, because that the kind states are only extends around the line defect. So as long as the Fermi level locates well inside the gap (not approaches the gap edges), the ribbon is too narrow and $\delta$ (model I) and $U$ (model II) is too small ($\delta$ or $U$ guarantees the formation and preservation of valley resolved kink states), the performance of the valley filters are well established.
Recently, by investigating the transport of electron wave packet through bilayer graphene quantum point contact, Costa {\it et al} found the system serves as a valley filter,\cite{Filter} in agreement with our numerical results.

\subsection{Monolayer graphene model with disorder}\label{PartII}

\begin{figure}
\includegraphics[width=\columnwidth, viewport=8 267 565 800, clip]{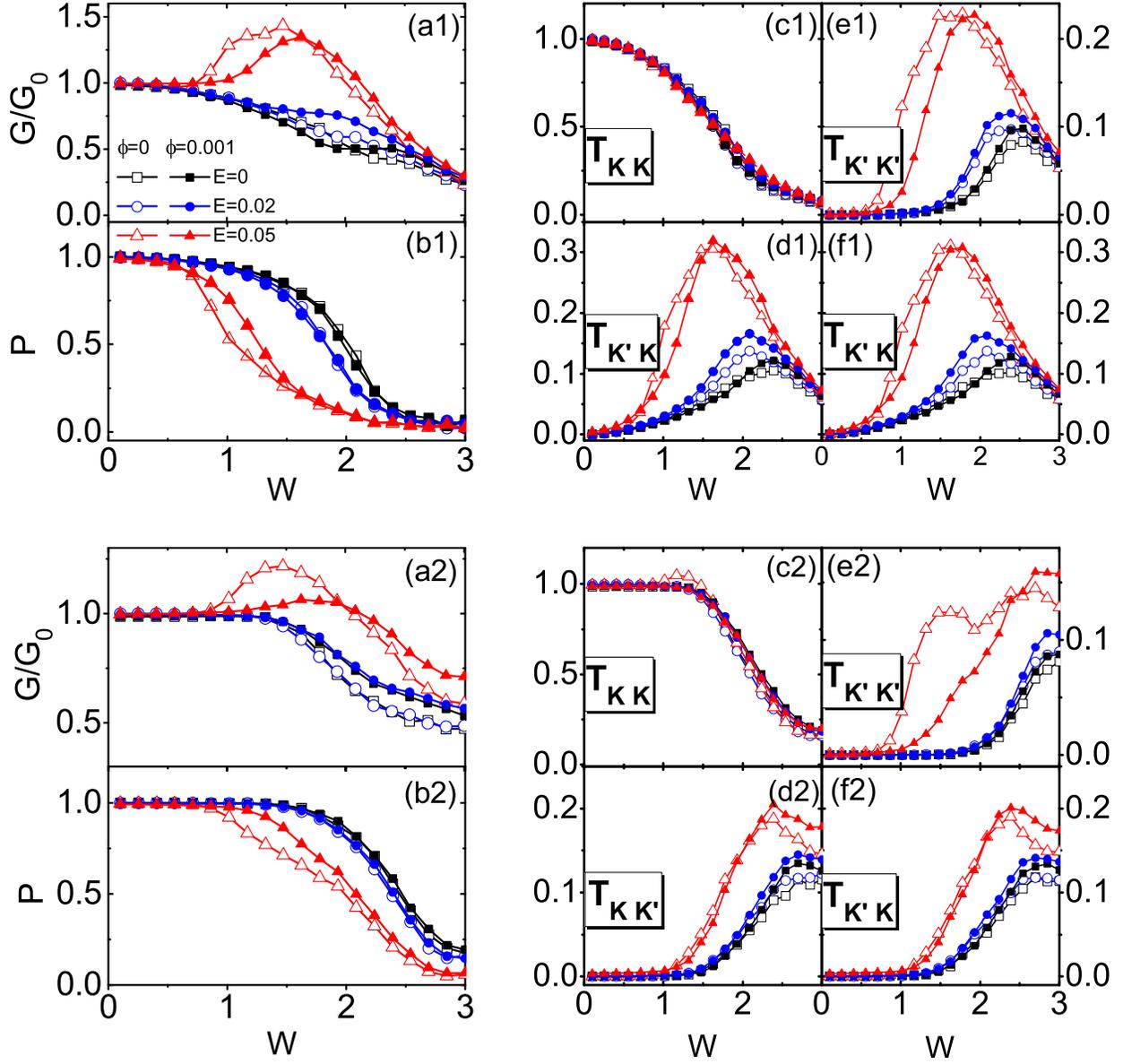}
\caption{(color online) The performance of model I in the presence of disorder. The results for Anderson (long range) disorder is exhibited in a1-f1 (a2-f2). The other parameters are $d=20$ and $M=100$. }\label{Fig3}
\end{figure}

In the following, to investigate the influence of disorder on model I, we set the line junction long enough (e.g. $M=100$, $\sim25\rm nm$) so that output carriers are fully polarized in the clean limit.
The results are shown in Fig. \ref{Fig3}. The cases for different Fermi level values and magnetic field are considered. For Anderson disorder, when the Fermi level $E$ is small (locates in the middle of band gap), the conductance $G$ is suppressed as the disorder strength $W$ is enhanced.
When $E$ approaches the band edge, $G$ increases from $G_0$ to form a peak in the diffusive regime and then drops fast for very large $W$.
Here the maximum value of the conductance $G$ is larger than $G_0$,
indicating that bulk states assisted intervalley transmission is rather strong.

The four valley resolved transmission coefficients $T_{K_2 K_1}$ are displayed in Fig. \ref{Fig3} (c1)-(f1). The only exist transmission in a clean sample, $T_{KK}$, is suppressed as the disorder strength $W$ increases and the decreasing trend is indifferent to $E$.
The other three scattering transmission coefficients (including $T_{K'K'}$, $T_{KK'}$ and $T_{K'K}$), are enhanced from $0$ to form a peak, and then are destroyed for very large $W$ for all cases.
When $E$ approaches the band edges, the scattering transmissions tend to happen for smaller $W$.
In this case, the bulk states in the central region are more easily to take part in the transmission and contributes to the conductance $G$.
So there are two types of competitive mechanisms dictate the total transmission: i) disorder damages the original permitted intravalley transmission $T_{KK}$, and ii) bulk states take part in the transmission and make a contribution to $G$.
The former mechanism happens in the whole process when electrons transmit from one end of the line junction to the other. The later mechanism, on the other hand, will be greatly influenced by the scattering at two ends of the line junction.
Take $K \rightarrow K'$ transmission as an example, incoming electron of valley $K$ from the left terminal, freely transmits into the line junction. The electron will be scattered by disorder as it passing through the whole region $\rm C$. At the interface of region $\rm C$ and $\rm R$, it is scattered into the $K'$ valley of the right terminal.
The counterpart process ($K' \rightarrow K$ transmission) has the similar experience, except that the intervalley scattering happens at the interface of region $\rm L$ and $\rm C$.
So a $K' \rightarrow K'$ scattering happens with at least twice intervalley scattering at both interfaces: $K' \rightarrow K\rightarrow K'$.
With above analysis, we can understand that the two intervalley transmission coefficients, $T_{KK'}$ and $T_{K'K}$, are almost the same.\cite{Valleydis} It is reasonable because the two events are spatially symmetric at two sides of region $\rm C$. Besides, the peak value of $T_{KK'}$ and $T_{K'K}$ is larger than $T_{K'K'}$.

In quantum Hall effect, magnetic field is applied in two dimensional gas to generate topological edge states. So how the valley resolved topological kink states are modulated by the magnetic field is an interesting question. In a very recent experimental work\cite{DomainBE1}, the transport of kink states under strong magnetic field is investigated. So next we focus on the effect of magnetic field on kink sates in graphene. When the magnetic field is applied, bulk electrons are tend to form Landau levels. While in the line junction, the kink states are only slightly separated.\cite{DomainBE2} So the effect of magnetic filed on transmission can be neglected when the Fermi level $E$ is small.
When $E$ is close to the band edge, the results are different. The discrepancy comes from two contrary mechanisms: i) the spacial departure between kink states and the bulk states (in the presence of magnetic field) weaken the scattering transmission; ii) bulk states under magnetic field are much localized against disorder, then scattering transmissions are better reserved.
So three scattering transmission coefficients are slightly suppressed by the magnetic field in the region of $W<2$.

Let us study the valley polarization $P$.
With the increase of the disorder strength $W$, the valley polarization $P$ decreases (see Fig. \ref{Fig3}(b1)). When $W$ is small, $P$ is quite robust against disorder.
For example, $P$ is about 0.9 still for small Fermi level $E$ while the disorder strength $W$ reaches 1.
For much stronger $W$, $P$ is weakened fast and then decays to nearly zero (e.g. $W=3$).
When the Fermi level $E$ moves close to the band edge,
the valley polarization $P$ becomes fragile against disorder and drops with smaller $W$.
This indicates that $P$ for smaller $E$ is much more robust against Anderson disorder.
This can also be ascribed to the participation of bulk states while $E$ locates in the vicinity of the band edge.
The bulk states, contains both many $K$ and $K'$ valley states, under the influence of disorder, participate in the transmission and then the conductance $G$ is strengthened and the valley polarization $P$ is weakened.
Similar to $G$, the valley polarization $P$ is only slightly affected by magnetic field.
Finally for very large $W$ all four components $T_{K_2 K_1}$ are almost the same and $P$ tends to zero.
So the results indicate that the decay of $G$ is mainly contributed by $T_{KK}$ for small $W$ and for very large $W$, though the four components are nonzero, equal values of $T_{K_2 K_1}$ leads to $P=0$.

The numerical results for long range disorder are shown in Fig. \ref{Fig3}(a2)-(f2).
Compared with the Anderson disorder cases, similar behaviors appears that for large Fermi level $E$,
the conductance $G$ is enhanced and the valley polarization $P$ is suppressed.
For the same set of parameters ($W$ and $E$), the magnetic filed only slight affects $G$ and $P$.
However, both the conductance $G$ and the valley polarization $P$ are more robust against long range disorder than Anderson disorder: both the conductance enhancement effect and $P$ decaying phenomena are weak.
When $E$ is small, $P=0.9$ can be achieved even for $W=2$. From Fig. \ref{Fig3}(c2), one tells that $T_{KK}$ is hardly affected by disorder and it begins decay until $W=1.5$ and meanwhile the other three components are slightly enhanced.
Compared with Anderson disorder, the disorder strength needed for scattering transmission is large. The results are quite reasonable since Anderson disorder may induce scattering between two states which have large discrepancy in the momentum space. However, long-range disorder only induces scattering between two states with small difference in momentum space.\cite{Scattering}

\subsection{Bilayer graphene model with disorder and two models' comparison}\label{PartIII}

\begin{figure}
\includegraphics[width=\columnwidth, viewport=8 267 565 800, clip]{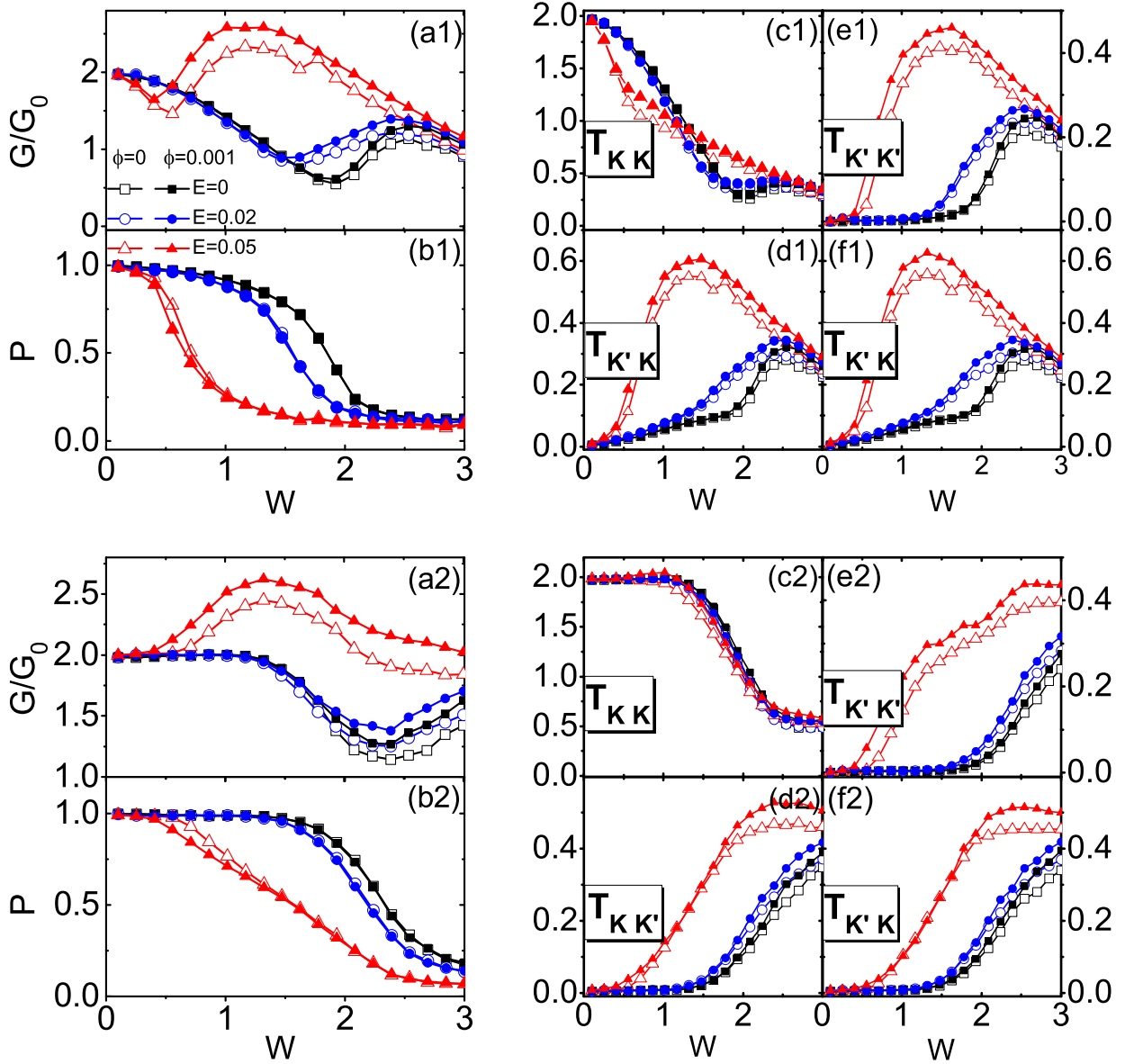}
\caption{(color online) $G$, $P$ and $T_{K_2K_1}$ versus $W$ for model II.
The results for Anderson (long range) disorder is exhibited in a1-f1 (a2-f2).
The size parameters are the same to Fig. \ref{Fig3}.
}\label{Fig4}
\end{figure}

Next we discuss the results for model II. In experiment, it is much easier to realized line junction based on bilayer graphene than monolayer graphene.\cite{DomainBE1,DomainBE2,DomainBE3}
The numerical results for model II with Anderson disorder (long range disorder) are exhibited in a1-f1 (a2-f2) of Fig. \ref{Fig4}.
The main results are similar to Fig. \ref{Fig3} for model I.
The conductance $G$ and the valley polarization $P$ in the present model under long range disorder preserve better quantized values than under Anderson disorder.
As the Fermi level $E$ moves toward the band edges, both $G$ and $P$ are much fragile against the same disorder strength.
Compared with model I, the disorder induced effects are much obvious. As Fermi level $E$ moves toward the band edges, there are several characters: i) $T_{KK}$ is weakened at small $W$; ii) scattering transmissions are active at smaller $W$; iii) $G$ is enhanced for large $W$ even $E$ is small.
On the other hand, while the Fermi level $E$ is not very close the band edges (e.g. $E=0$ and $0.02$),
the valley polarization $P$ is quite robust against the disorder. For example, $P$ still can reach $0.9$ at the disorder strength $W=1$ for the Anderson disorder case [see Fig. \ref{Fig4}(b1)], and
$P$ can almost keep $1$ at $W=1$ under the long range disorder regardless of the magnetic field
[see Fig. \ref{Fig4}(b2)].

\begin{figure}
\includegraphics[width=\columnwidth, viewport=58 272 505 804, clip]{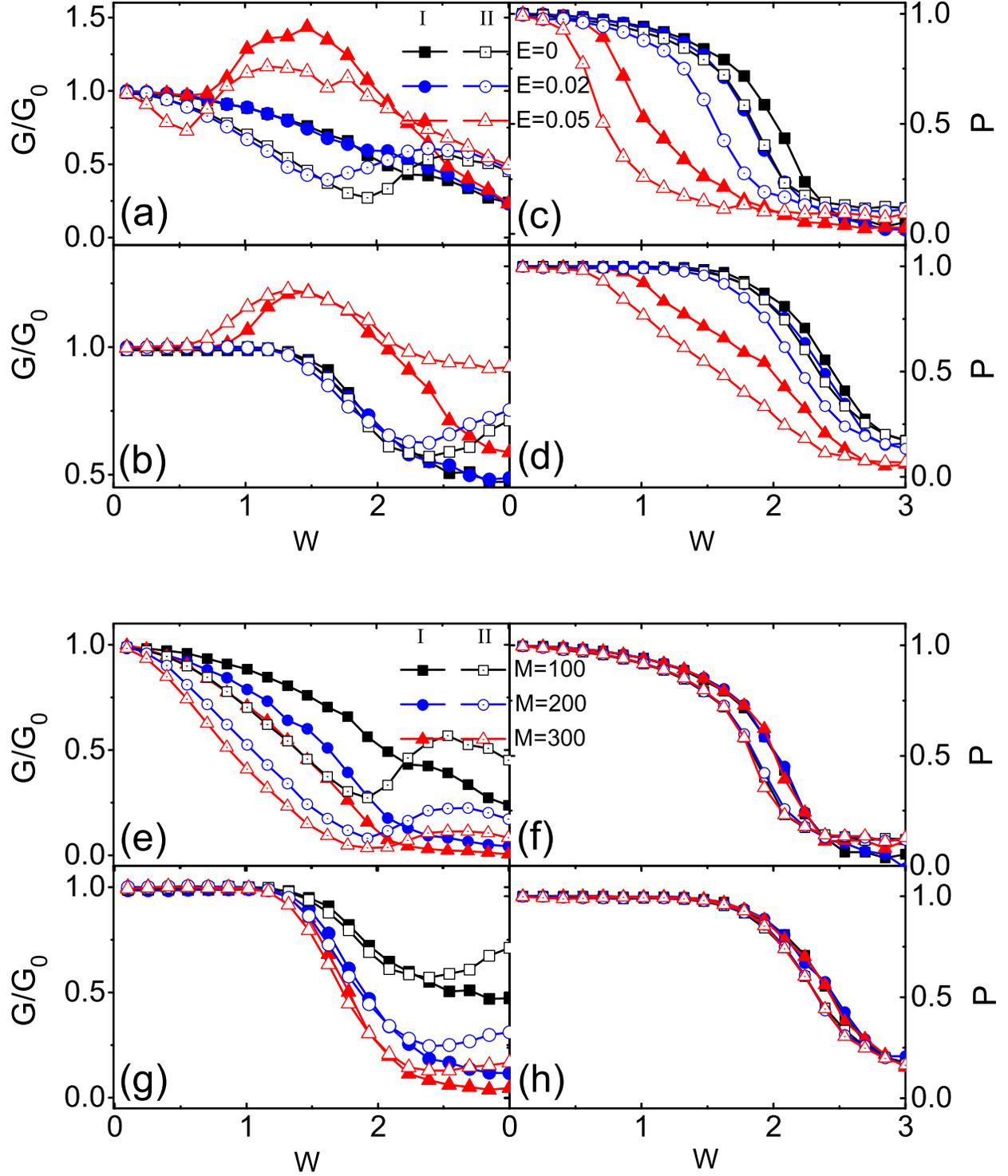}
\caption{(color online) The performances comparison between model I and II for different $E$ (upper panel) and different $M$ (lower panel). The result for Anderson disorder (a, c, e ,h) and long range disorder (b, d ,f, h) are shown, respectively. In the upper panel, we use $M=100$ and in the lower panel, $E=0$ is adopted. In this figure, $\phi=0$.
}\label{Fig5}
\end{figure}

As well known the carriers in monolayer graphene are massless Dirac fermions and in bilayer graphene they are massive.
Klein theory predicted that massless Dirac Fermions can tunneling through a potential barrier freely when normally injected.\cite{Klein}
So we wonder, in the present two models, which one is better served as valley filter.
The parameters for both models are intentionally selected that the band gaps are the same $\sim[-0.061,0.061]$. The numerical results are shown in Fig. \ref{Fig5}.
The solid and hollow symbols represent results for model I and II, respectively.
Here the conductance $G$ for model II is divided by $2$ thus the comparison can be made between two models.

Under both types of disorder, when the Fermi level $E$ is small,
$G$ versus $W$ decrease monotonously for both models due to the backscattering effect.
As $E$ increases, $G$ for both models show similar behaviors.
However, the valley polarization $P$ versus $W$ for model II decays fast than curves for model I.
The results indicate that in the present models, valley polarization $P$ is better preserved in monolayer graphene line defect.
However, $P$ is very large still for both models under the moderate disorder strength (e.g. $W=1$) as long as the Fermi level $E$ is not very close to the gap edge.
While $E$ being near the gap edge, there are more bulk states in model II, so the scattering is stronger, and leading the quick decay of $P$ versus $W$.
To confirm this speculation, we also check the results for model II with larger $N$ (numerical results not shown). When $N$ becomes large, there are more bulk states near the gap edges and the band gap stay unchanged.
Similar results are obtained that in the diffusive regime, $G$ is enhanced stronger and $P$ is smaller.

\subsection{Length dependent relations}\label{PartIV}

In above discussion, we have assumed the length of the line junction is $M=100$ in both models. Under weak disorder, this length is long enough to serve as a perfect valley filter. As the length of line defect becomes longer, it is reasonable to speculate that the backscattering should be enhanced. The scattering transmissions are affected in two ways: intervalley scattering happens at the ends of the line junction and backscattering happen during the propagation along the line junction. The former one, is independent of the line junction length, as long as the length is long enough (e.g. the length $M>100$). The later one is suppressed for larger $M$. In Fig. \ref{Fig5} (e)-(h), the conductance $G$ and the valley polarization $P$ versus $W$ relations for different length $M$ are shown with both types of disorder considered with $E=0$. For Anderson disorder, when $M$ is larger, $G$ decreases faster from (scaled) unit value. But for large $W$, $G$ is enhanced for bilayer model. It is because in model II there are much more bulk states near the gap and can aid the transport in the strong disorder.

One interesting finding is that the curves of the valley polarization $P$ for different length $M$ coincide with each other [see Fig. \ref{Fig5}(f) and (h)].
This indicates that $P$ are hardly affected by the length $M$ as long as $M$ is not very short,
which is similar with the results of the clean sample as shown in Fig. \ref{Fig2}(c) and (d).
So for a valley filter long enough, the valley polarization $P$ is only determined by the disorder strength even through the transmission is suppressed heavily.
Under a specific disorder, when the line defect becomes longer, all valley resolved coefficients (including intervalley scattering coefficients $T_{KK}$, $T_{K'K'}$ and intravalley scattering coefficients $T_{KK'}$, $T_{K'K}$) are suppressed at the same rate, thus $P$ is unchanged following Eq. (\ref{EQ4}).
In experiment, when the line junction is subjected to diffusive regime and its length is longer than the mean free path, the valley polarization can always well preserved only if the disorder is not too strong.
(f) and (h) in Fig. \ref{Fig5} show analogous characters except that the valley polarization $P$ are much more robust against long range disorder than Anderson disorder.
When $E$ is fixed and $M$ becomes larger, the scattering transmission induced conductance enhanced effect is weakened and then $G$ decreases monotonously with the increase of $W$.

\begin{figure}
\includegraphics[width=\columnwidth, viewport=22 32 757 546,clip]{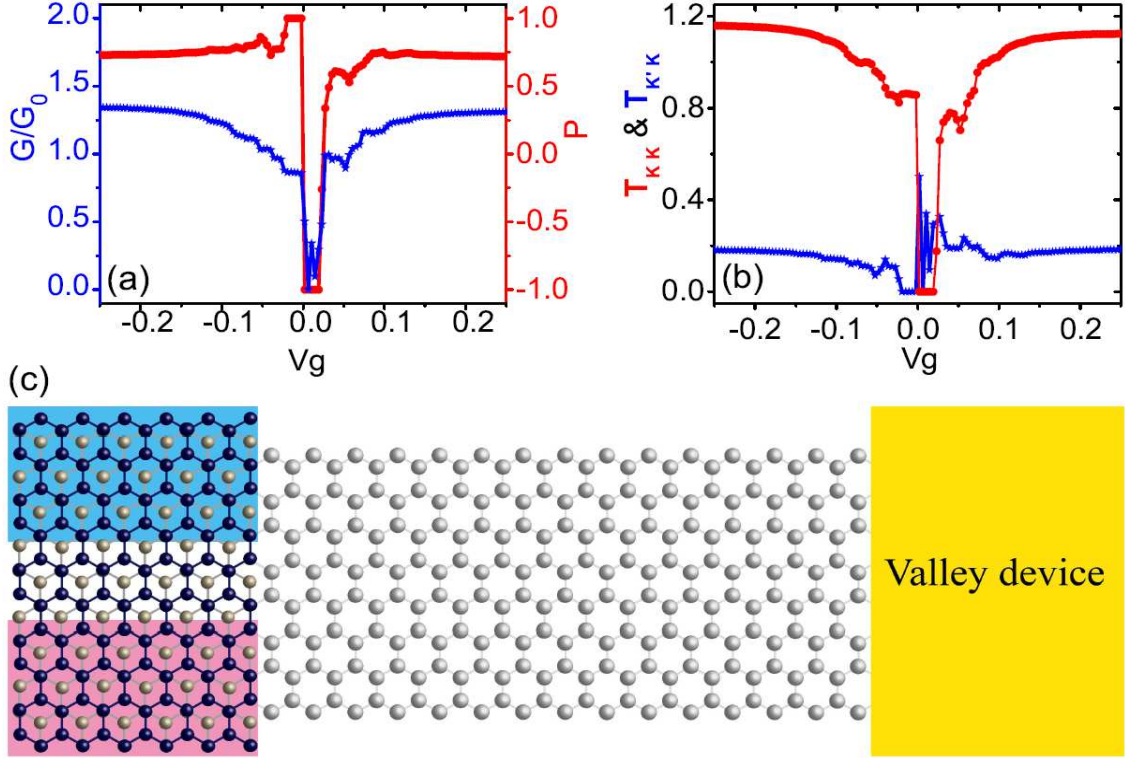}
\caption{(color online) Characters of valley polarized current injection from bilayer graphene line defect to normal monolayer graphene. (a) $G$ and $P$ versus $V_g$. (b) $T_{KK}$ and $T_{K'K}$ versus $V_g$. (c) The schematic of hybrid model composed of a bilayer graphene based line defect terminal, normal monolayer graphene nanoribbon and a valley device.
}\label{Fig6}
\end{figure}

\subsection{Hybrid graphene model}\label{PartV}

To generate valley polarized current with monolayer graphene, one needs to align monolayer graphene to hexagonal boron nitride substrate and form a topological line defect. Up to now, it has not be observed in experiment.  Instead, experiment physicists have successfully realized the quasi one-dimensional line defect in bilayer graphene,\cite{DomainBE1,DomainBE2,DomainBE3} so it is favored to generate valley polarized current based on bilayer graphene. Once the valley polarized current is realized we need to transport it for application. Usually monolayer graphene bears much higher mobility than the bilayer graphene samples.\cite{Review1} So in valleytronics devices, it is favored to transport valley polarized current with monolayer graphene for application as shown in Fig. \ref{Fig6} (c). Here we discuss a hybrid junction with the left side a bilayer graphene line defect based semi-infinite terminal and the right side a normal monolayer graphene. Similarly, to eliminate the effect of edge states, the left terminal is wider than the right terminal ($N=200$ for the left ribbon and $N=120$ for the right one). So the right going injection electron is initially valley polarized (valley index $K$ for instance).

The numerical results are shown in Fig. \ref{Fig6} (a) and (b). We change the gate voltage of the right terminal $V_g$ to investigate the behaviors of the conductance $G$, valley polarization $P$ and valley resolved transmission coefficients $T_{KK}$ and $T_{K'K}$ ($T_{KK'}$ and $T_{K'K'}$ are infinitesimal).
The conductance $G$ is generally smaller than $2G_0$ but larger than $G_0$ except the near neutral point region that $G$ shows a dip.
Meanwhile, $P$ is around $0.75$ when $V_g$ changes except in a narrow region around the neutral point, $P=-1$.
For $V_g$ being not very close zero, the transmission mode towards the positive direction at $K$ in monolayer graphene is always available, leading a large intravalley transmission coefficient $T_{KK}$
and a small intervalley scattering transmission coefficient $T_{K'K}$.
In this case, as shown in Fig. \ref{Fig6}(b), $T_{KK}$ is above $1.0 G_0$ always, and $T_{K'K}$
is around $0.2G_0$ which is much smaller than $T_{KK}$.
So the valley polarization is still quite large.
On the other hand, while in the vicinity of neutral point, only several modes are available.
In particular, as $V_g$ being a small positive values, the available transmission modes locates at $K'$, thus carrier flowing from $K$ of the left bilayer graphene to $K$ of the right monolayer one is completely blocked and $T_{KK}$ tends to zero.
Then $T_{KK'}$ is dominant which leads $P=-1$ in a rather small interval.

\section{Discussion and Conclusion}\label{conclusion}
The valley filter was first proposed by Rycerz  {et al},\cite{Beenakker} which is composed of a narrow graphene nanoribbon sandwiched by two ribbons of large width. Similar to line defect models, the propagation directions of the states are associated with the valley indexes and valley polarized current are generated with a voltage bias. In Rycerz's model, such states is realized by narrow edges states located at the edges of the ribbon. So the transport of the states is sensitive to roughness of the ribbon’s edges and hence ruins the valley filter efficiency. In the graphene defect line model, however, topological kink states are well localized around the line defect \cite{Domain4}. The channels of such state are rather wide (determine by max$\{\hbar v_F/\delta, d\}$ with $v_F$ the Fermi velocity). Thus the forward valley polarized kink state cannot be easily backscattered\cite{Domain4}.

In conclusion, we have investigated the transport properties of line defect valley filter based on monolayer and bilayer graphene model.
In the ballistic regime, when the length of the line defect is larger than $15 \rm nm$, the output electrons are almost $100\%$ valley polarized and the conductance is quantized.
While in the moderate disorder strength regime(e.g. $W=1t$), the valley filter can well survive and its valley polarization efficient can reach $90\%$ still except for the Fermi level approaches the band edge. While under the strong disorder, the system subjects to diffusive regime.
In this case, the transport is mainly affected by interface backscattering
and bulk states assisted scattering transmission,
leading to the enhancement of conductance and suppression of valley polarization.
At last, while disorder is very strong, the conductance is reduced again and the valley polarization is almost zero.
Under the same disorder, as the line defect becomes longer, the conductance decays but the valley polarization efficient does not change.
This character is quite useful that we can manipulate electron valley in a longer line defect with no expense of valley polarization.
In experiment the line junction valley filter is easier to be realized by engineering bilayer graphene. So we discuss the valley injection efficiency in a hybrid system composed of bilayer graphene line defect and normal monolayer graphene. The results show that the valley can efficiently be injected
from the bilayer graphene valley filter to the monolayer graphene.

\section*{ACKNOWLEDGMENTS}
We gratefully acknowledge fruitful discussions with Prof. X. C. Xie and Prof. H. W. Liu.
This work was financially supported by NBRP of China (Grants Nos. 2014CB920901, 2015CB921102 and 2012CB921303) and NSF-China under Grants Nos. 11534001, 11374219, 11274364 and 11574007.

\end{document}